\documentclass{aa}
\usepackage{graphicx}
\usepackage{times}

\begin{document}

\def\grad{\hbox{$^\circ$}}
\def\bsec{\hbox{$^{\prime\prime}$}}
\def\std{\hbox{$^{\rm h}$}}
\def\min{\hbox{$^{\rm m}$}}
\def\sec{\hbox{$^{\rm s}$}}
\def\mag{\mbox{ mag}}

\thesaurus{05
           (08.08.1;         
            11.19.4;         
            11.13.1)         
           }

\title{Studies of binary star cluster candidates in the bar of the LMC. II.
\thanks{Based on observations taken at the European Southern Observatory, La
  Silla, Chile in program 54.D-0730 and during time allocated by the MPIA,
  Heidelberg.}} 

\author{Andrea Dieball\inst{1}, Eva K. Grebel\thanks{Hubble Fellow}\inst{2}}

\institute{Sternwarte der Universit\"at Bonn, Auf dem H\"ugel 71, 
           D--53121 Bonn, F.R. Germany \and
           University of Washington, Department of Astronomy, Box 351580, 
           Seattle, WA 98195-1580, USA }

\offprints{Andrea Dieball, adieball@astro.uni-bonn.de}

\date{Received 3 August 1999, accepted 30 March 2000}

\maketitle

\markboth{A. Dieball \& E. K. Grebel: Binary clusters in the LMC bar. II.}{A. Dieball \& E. K. Grebel: Binary clusters in the LMC bar. II.}

\begin{abstract}

Binary clusters account for more than 10\% of the cluster population in the
Magellanic Clouds.  Statistically fewer than 50\% of the found pairs are
expected to be chance superpositions. We estimated the cluster encounter rate
and suggest that tidal capture is an unlikely formation scenario for the
formation of binary clusters. Thus, most {\it true} binary clusters can be
expected to have formed together.    

Here we present a study of three binary cluster candidates which are located in
the bar of the LMC.  
NGC 1971 \& NGC 1972 are situated in the association LH\,59 in the eastern
part of the bar. A third star cluster, NGC 1969, is close enough to this pair
that all three objects may constitute a triple system. We present the first age
determination that is based on CMDs for these star clusters. Our findings
suggest that all three clusters are young (40-70 Myr) and may have been formed
in the same GMC. It cannot clearly decided whether the clusters are physically
interacting or not. 

NGC\,1894 \& SL\,341 are located at the south-western rim of the LMC bar. This
pair is studied in detail for the first time: The isopleths of both clusters
reveal an elliptical shape. Whether this might be interpreted as a sign of
interaction or is a peculiarity which is shared with a large amount of LMC
star clusters which show higher ellipticities than their counterparts in the
Milky Way remains unclear. From our age determination we find that both
clusters are coeval with an age of $55\pm5$ Myr. This makes a formation from
the same GMC a likely scenario. 

SL\,385 \& SL\,387 are a close pair in the western part of the LMC bar. We
derived ages of $170\pm30$ Myr for SL\,385 and $\ge250$ for SL\,387. The large
age difference makes it unlikely that these two clusters formed in the same
GMC. 

\keywords{Magellanic Clouds -- Hertzsprung-Russel (HR) and C-M diagrams --
          Galaxies: star clusters: NGC 1971, NGC 1972, NGC 1969 --
          Galaxies: star clusters: SL\,385 \& SL\,387 -- Galaxies: star
          clusters: NGC\,1894 \& SL\,341}   

\end{abstract}

\section{Introduction}
\label{intro}

The Magellanic Clouds offer the unique possibility to study star clusters in
general and binary clusters in particular. These two companion galaxies are 
close enough to resolve single stars, but distant enough to make the detection
of close pairs of star clusters an easy task. 
Bhatia \& Hatzidimitriou (\cite{bh}), Hatzidimitriou \& Bhatia (\cite{hb}),
and Bhatia et al. (\cite{brht}) have surveyed the Magellanic Clouds in order
to catalogue the binary cluster candidates. The maximum projected
centre-to-centre separation of the components of a pair chosen for inclusion
in the list of candidates was 18 pc, which corresponds to $\approx 1\farcm3$
in the Large Magellanic Cloud (LMC). A binary cluster with larger separation
may become detached by the external tidal forces while shorter separations may
lead to mergers (Sugimoto \& Makino \cite{sm}, Bhatia \cite{bhatia}).     
Two clusters may appear to be a binary cluster due to chance line-up while 
in fact being at different distances within the Magellanic Clouds and not
gravitationally bound to each other. An estimation of the number of such 
chance-pairs of objects (Page \cite{page}) revealed that less than half of all
pairs found are expected statistically. As considerably more
double clusters were found, this strongly suggests that at least a
subset of them must be true binary clusters, i.e., clusters that are formed
together and/or may interact or even be gravitationally bound. 

Star clusters form in giant molecular clouds (GMCs), but the details of
cluster formation are not yet totally understood (Elme\-green et
al. \cite{eepz}). Fujimoto \& Kumai (\cite{fk}) suggest that star clusters form
through strong collisions between massive gas clouds in high-velocity random
motion. Oblique cloud\--cloud collisions result in compressed sub-clouds which
revolve around each other and form binary or multiple clusters. Binary star
clusters are expected to form more easily in galaxies like the Magellanic
Clouds, whereas in the Milky Way the required large-scale high-velocity random
motions are lacking. Indeed only few binary clusters are known in our own
Galaxy.  

The probability of tidal capture of one cluster by another one is small
(Bhatia et al. \cite{brht}), but becomes more probable in dwarf galaxies like
the Magellanic Clouds with small velocity dispersion of the cluster systems
(van den Bergh \cite{vdbergh}). In that case the clusters would be
gravitationally bound, but age differences are likely. Tidal interactions
between clusters can be traced using isodensity maps (de\,Oliveira et
al. \cite{dodb}, Leon et al. \cite{lbv}). However, 
Vallenari et al. (\cite{vbc}) estimated a cluster encounter rate of
$dN/dt\sim1\cdot(10^{9} \mbox{yr})^{-1}$. This makes tidal capture of young
clusters very unlikely. Interaction between LMC and SMC might have led to a
higher cluster formation rate and, due to the dynamical perturbance, a higher
encounter rate might be possible (see Vallenari et al. \cite{vbc} and
references therein).   
      
The formation of low-mass star clusters tends to proceed hierarchically in
large molecular complexes over several $10^7$ years (e.g., Efremov \& Elmegreen
\cite{ee}). Leon et al. (\cite{lbv}) suggest that in these groups the cluster
encounter rate is higher and thus tidal capture is more likely: Binary 
clusters are not born together as a pair with similar ages but are formed
later through gravitational capture. An observational test of this scenario
would require the detection of evidence of tidal interactions between
clusters, whose age differences need to be compatible with the survival times
of GMCs. 

Another binary cluster formation scenario is introduced by Theis et
al. (\cite{teph}) and Ehlerov\'a et al. (\cite{epth}): Exploding supernovae
close to the centre of a giant molecular cloud sweep up the
outer cloud material within a few Myrs and accumulate it in the shell. The
large amount of matter makes the shell prone to gravitational fragmentation
and finally to the formation of many open cluster-like associations. In case
of a dense ambient medium outside the cloud or a very massive original
molecular cloud the shell can be strongly decelerated resulting in a
recollapse. A galactic tidal field acting on this recollapsing shell can split
it into two or more large clusters, thus forming coeval twin clusters which
are, however, not gravitationally bound (Theis \cite{theis}). The fragments
may stay together for a long time, though they are gravitationally
unbound. The evolution of their spatial separation mainly depends on the shape
of the shock front at the time of fragmentation. 

Depending on their masses and separations binary clusters will eventually merge
or become detached. A merged binary cluster will have one single but 
elliptical core (Bhatia \& McGillivray \cite{bm}) which can also be found in
the young blue populous star clusters of the Magellanic Clouds. Can mergers of
former binary star clusters be responsible for at least some of the blue
populous clusters in the LMC?   

We are studying binary cluster candidates in the Magellanic Clouds to
investigate whether the cluster pairs are of common origin and if they show
evidence for interaction. While it is not possible to measure true,
deprojected distances between apparent binary clusters an analysis of their
properties can give clues to a possible common origin. 

In this paper, we investigate three binary cluster candidates, namely
NGC\,1894 (also known as SL\,344, Shapley \& Lindsay \cite{sl}, or BRHT\,8a,
see Bhatia et al. \cite{brht}) \& SL\,341 (or BRHT\,8b), NGC\,1971 \&
NGC\,1972 (or SL\,481 (BRHT\,12a) \& SL\,480 (BRHT\,12b), and SL\,387 
\& SL\,389 (or BRHT\,35a \& BRHT\,35b). All three cluster pairs are 
located in the bar of the LMC.  

NGC\,1894 and SL\,341  constitute a close pair at the south-western rim of
the LMC bar. Only the bigger cluster NGC\,1894 appears in the list of Bica et
al. (\cite{bcdsp}) as an SWB type II cluster (Searl et al. \cite{swb}). Here
we present the first detailed study of this candidate binary cluster. 

NGC\,1971 and NGC\,1972 are situated in the association
(or stellar cloud, Lucke \& Hodge \cite{lh}) LH\,59 in the eastern part of the
LMC bar, close to the geometrical centre of this galaxy (see Hodge \& Wright
\cite{hw}).  A third cluster, NGC\,1969 (or SL\,479), is located within
$1\farcm5$ ($\approx22$ pc). Bica et al. (\cite{bcdsp}) classified the cluster
pair to be of SWB type II while the third cluster belongs to the next older
class SWB type III. So far, no accurate ages based on colour magnitude
diagrams (CMDs) have been obtained for this potential triple cluster system. 

The third cluster pair, SL\,385 and SL\,387, is also located close to the
geometrical centre, but in the western part of the LMC bar.   
According to ages for a large sample of star clusters, derived on the base of
integrated colours, Bica et al. (\cite{bcdsp}) propose an age gradient of the
LMC bar. Younger clusters are predominantly found in the eastern part, while
older clusters of SWB type III and higher are concentrated to its western
end. From integrated colours Bica et al. (\cite{bcdsp}) find these two clusters
to be coeval, while Vallenari et al.\ (1998) determine a large CMD-based age
difference of $\sim 350$ Myr between the clusters. Based on subgiant density
profiles Leon et al. (\cite{lbv}) suggest that the two clusters may constitute
a true binary cluster with physical interaction.       

This paper is organized as follows. In Sect. \ref{phot} we describe the
photometric data in general. Stellar density maps are presented in
Sect. \ref{stardens}. The following section describes the CMDs for the
components of each cluster pair. Ages for each cluster are derived and
compared with previous studies. In Sect. \ref{tidalcapture} we estimate the
probability of cluster encounters in the LMC's bar and disk. In
Sect. \ref{summary} we give a summary and discuss the results.  

\section{Photometric observations and data reduction}
\label{phot}

The cluster pair NGC\,1894 \& SL\,341 was observed on February 8, 1995, with
ESO Multi-Mode Instrument (EMMI) using the red arm (Red Imaging and Low
dispersion spectroscopy -- RILD) at the ESO New Technology Telescope (NTT) at
La Silla. The ESO \#36 chip (TEK $2048\times2048$) was used with a pixel scale
of $0\farcs27$. The resulting field of view is $9\farcm2\times8\farcm6$. The
data were obtained with the Bessell $B$, $V$, $R$ and Gunn $i$-filters.
 
The data for the other two binary cluster candidates were obtained on March 23
(NGC 1971 \& NGC 1972), and 27 (SL 385 \& SL 389), 1994, with the ESO
Faint Object Spectrograph and Camera 2 (EFOSC 2) at the ESO/MPI 2.2 m
telescope at La Silla. A $1024 \times 1024$ coated Thomson 31156 chip (ESO
\#19) was used with a pixel scale of $0\farcs 34$ resulting in a field of view
of $5\farcm8 \times 5\farcm8$.  
These data were obtained with the Washington $T1$, Gunn $g$ (which resembles
Washington $M$), and Gunn $i$ (which corresponds to Washington $T2$) filters
used at the 2.2 m telescope. In order to avoid flatfielding problems near the
edges of the exposures taken in Washington-filters, each image was cut out
resulting in a field of view of $4\farcm3 \times 3\farcm5$. The problem is
worst in $T1$ and least in Gunn $i$. Images of the binary cluster candidates
are shown in Figs. \ref{n1971ps}, \ref{sl385ps}, and \ref{n1894ps}.  

Table \ref{obs} gives an observing log.

After standard image reduction with MIDAS, profile fitting photometry
was carried out with DAOPHOT~II (Stetson \cite{stetson}) running under MIDAS.

The photometry was transformed using standard fields observed in the same
nights the object data were obtained. For the Washington data the standard
fields SA\,110 and SA\,98 (Geisler \cite{geisler}), and for the Bessell data
the standard fields around Rubin\,149, PG\,1942, PG\,0942 and the SA\,98
(Landolt \cite{landolt}) were used.

We applied the following transformation relations:
\begin{eqnarray*}
&&g-G = z_{G} + a_{G} \cdot X + c_{G} \cdot (G-T1)\\
&&t1-T1 = z_{T1} + a_{T1} \cdot X + c_{T1} \cdot (G-T1)\\
&&i-I = z_{I} + a_{I} \cdot X + c_{I} \cdot (T1-I)\\
&&b-B = z_{B} + a_{B} \cdot X + c_{B} \cdot (B-V)\\
&&v-V = z_{V} + a_{V} \cdot X + c_{V} \cdot (B-V)\\
&&r-R = z_{R} + a_{R} \cdot X + c_{R} \cdot (V-R)\\
&&i-I = z_{I} + a_{I} \cdot X + c_{I} \cdot (V-I),
\end{eqnarray*}
where $X$ is the mean airmass during observation, capital letters
represent standard magnitudes and colours, and lower-case
letters denote instrumental magnitudes after normalizing to
an exposure time of 1 sec. The resulting colour terms $c_{i}$, zero points
$z_{i}$ and atmospheric extinction coefficients $a_{i}$ are listed in Table \ref{transf} for all nights.

\begin{table}
\caption[]{\label{obs}Observing log}
\begin{tabular}{lcrc}
\hline
Object             & Filter    &\multicolumn{1}{c}{Exp.time}& Seeing  \\
                   &           & [sec]                      &[\arcsec]\\
\hline
NGC\,1971/NGC\,1972 & Gunn $g$ & 600                        & 1.2 \\
NGC\,1971/NGC\,1972 & Gunn $g$ & 60                         & 1.4 \\
NGC\,1971/NGC\,1972 & $T1$     & 300                        & 1.3 \\
NGC\,1971/NGC\,1972 & $T1$     & 30                         & 1.3 \\
NGC\,1971/NGC\,1972 & Gunn $i$ & 300                        & 1.1 \\
NGC\,1971/NGC\,1972 & Gunn $i$ & 30                         & 1.1 \\ \hline
SL\,385/SL\,389     & Gunn $g$ & 600                        & 1.3 \\
SL\,385/SL\,389     & Gunn $g$ & 60                         & 1.5 \\
SL\,385/SL\,389     & $T1$     & 240                        & 1.2 \\
SL\,385/SL\,389     & $T1$     & 30                         & 1.1 \\
SL\,385/SL\,389     & Gunn $i$ & 240                        & 1.4 \\
SL\,385/SL\,389     & Gunn $i$ & 30                         & 1.5 \\
SL\,385/SL\,389     & Gunn $i$ & 30                         & 1.2 \\ \hline
NGC\,1894/SL\,341   & $B$      & 300                        & 1.2 \\
NGC\,1894/SL\,341   & $B$      & 40                         & 1.3 \\
NGC\,1894/SL\,341   & $V$      & 130                        & 1.3 \\
NGC\,1894/SL\,341   & $V$      & 10                         & 1.3 \\
NGC\,1894/SL\,341   & $R$      & 180                        & 1.0 \\
NGC\,1894/SL\,341   & $R$      & 10                         & 0.9 \\
NGC\,1894/SL\,341   & Gunn $i$ & 180                        & 1.0 \\
NGC\,1894/SL\,341   & Gunn $i$ & 10                         & 0.9 \\ \hline
\end{tabular}
\end{table}

\begin{table}
\caption[]{\label{transf}Transformation coefficients}
\begin{tabular}{lrcc}
\hline
Filter   & \multicolumn{1}{c}{$c_{i}$} & $z_{i}$ &  $a_{i}$\\
         &         & [mag]   & [mag]   \\
\hline
\multicolumn{4}{c}{March 23, 1994}\\ \hline
Gunn $g$ & $-0.108\pm0.008$ & $1.753\pm0.020$ & $0.147\pm0.012$ \\
$T1$     & $ 0.093\pm0.010$ & $2.569\pm0.025$ & $0.118\pm0.015$ \\
Gunn $i$ & $ 0.126\pm0.005$ & $2.414\pm0.010$ & $0.085\pm0.006$ \\ \hline
\multicolumn{4}{c}{March 27, 1994}\\ \hline
Gunn $g$ & $-0.093\pm0.005$ & $1.713\pm0.011$ & $0.145\pm0.006$ \\   
$T1$     & $ 0.092\pm0.004$ & $2.614\pm0.008$ & $0.096\pm0.004$ \\   
Gunn $i$ & $ 0.115\pm0.006$ & $2.475\pm0.009$ & $0.046\pm0.005$ \\ \hline
\multicolumn{4}{c}{February 8, 1995}\\ \hline
$B$      & $ 0.032\pm0.022$ & $1.863\pm0.028$ & $0.215\pm0.016$ \\   
$V$      & $-0.032\pm0.016$ & $1.253\pm0.021$ & $0.121\pm0.010$ \\   
$R$      & $ 0.007\pm0.028$ & $1.109\pm0.021$ & $0.076\pm0.009$ \\   
Gunn $i$ & $ 0.029\pm0.012$ & $1.752\pm0.013$ & $0.046\pm0.006$ \\ 
\hline
\end{tabular}
\end{table} 

\section{Stellar density maps}
\label{stardens}

\begin{figure}
\centerline{
\includegraphics[width=8.7cm]{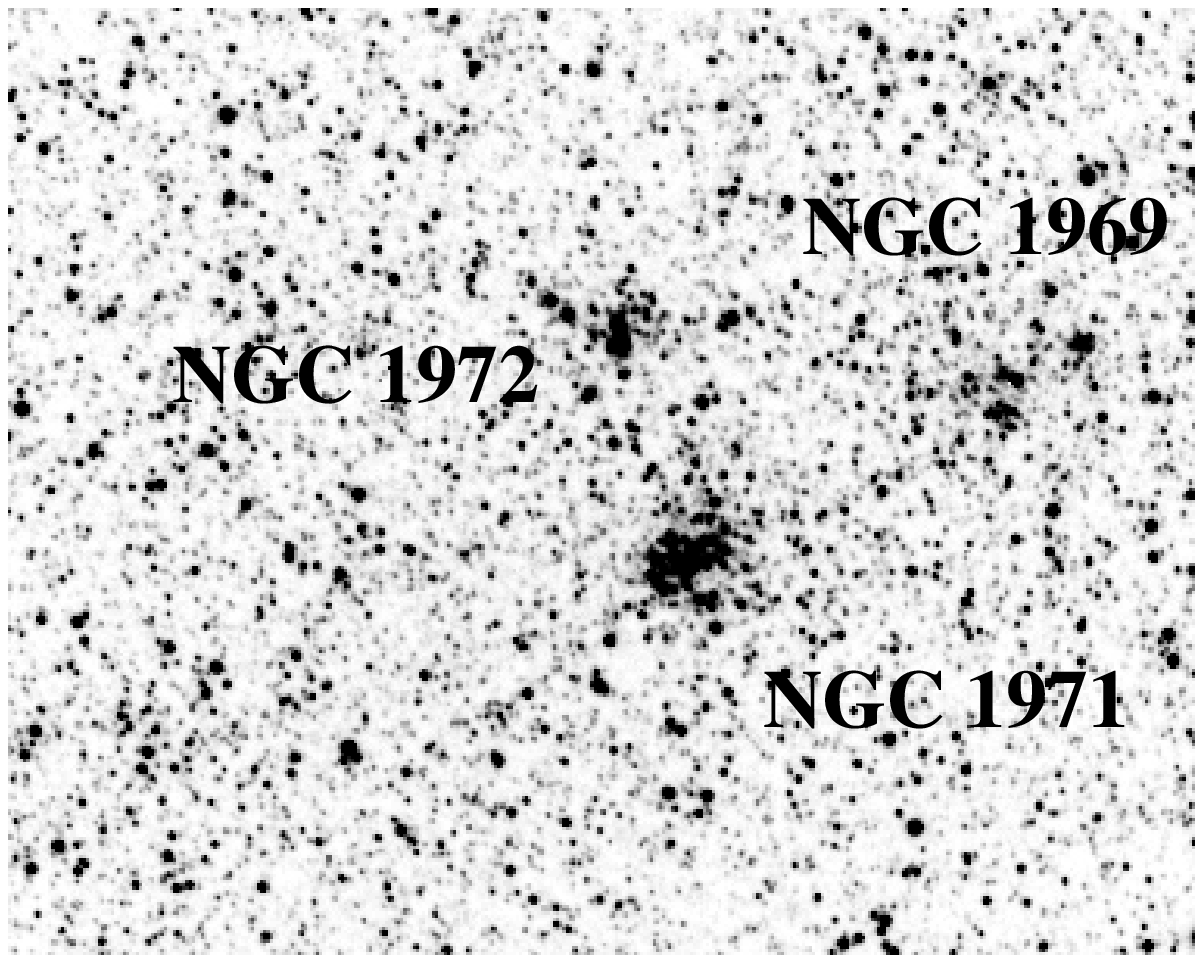}
}
\caption[]{\label{n1971ps} T1-image of the cluster pair NGC\,1971 \&
  NGC\,1972. North is up and east to the left. The field of view is $4\farcm3
  \times 3\farcm5$. This candidate binary cluster is located in the bar of the
  LMC close to the geometrical centre. To the northwest a third close cluster
  is located: NGC\,1969} 
\vspace{0.5cm}
\centerline{
\includegraphics[width=8.7cm]{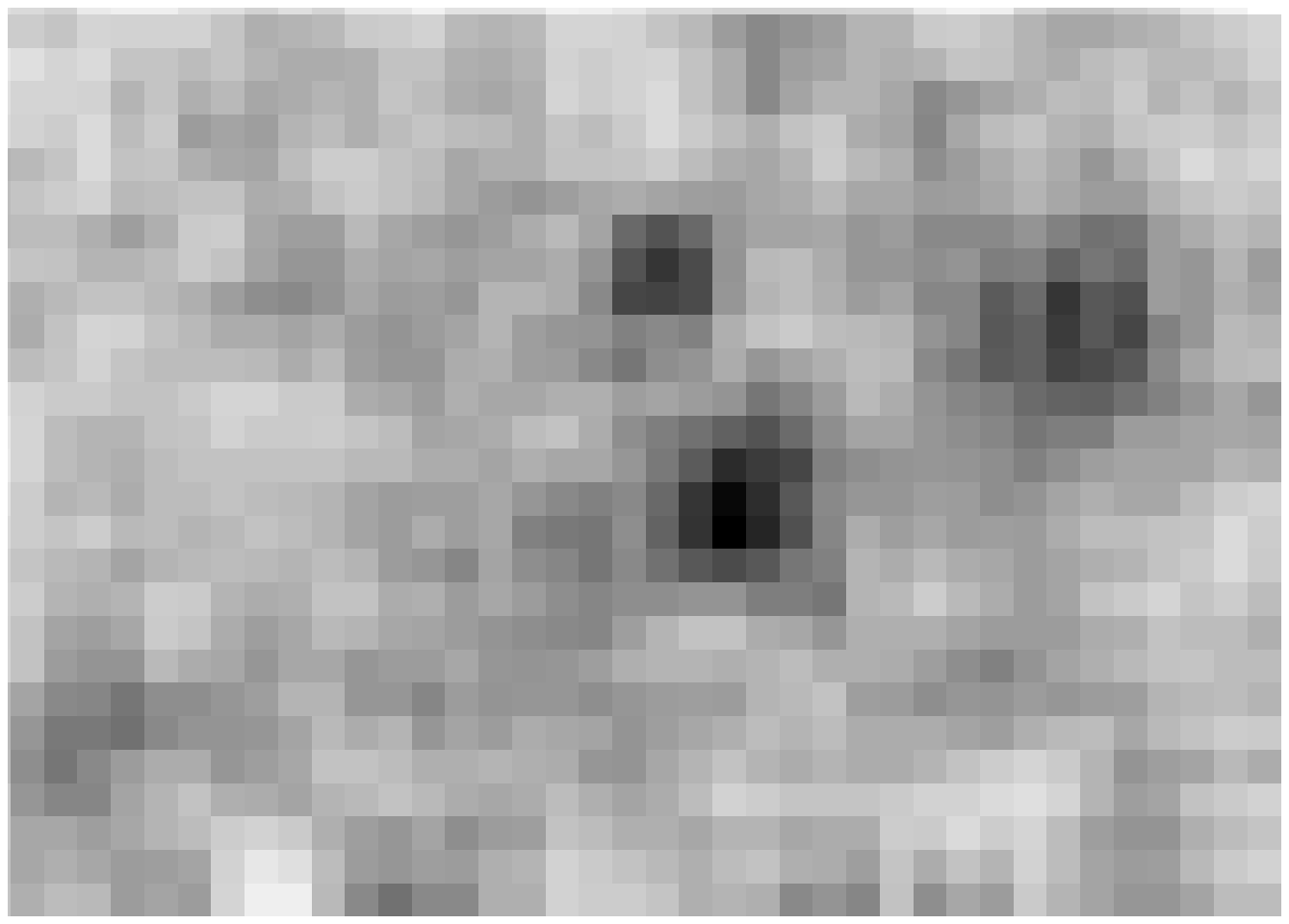}
}
\caption[]{\label{n1971dens} Stellar density in and around the three clusters
  NGC\,1971, NGC\,1972 and NGC\,1969. One pixel of the density map corresponds
  to $20\times20$ pixels in the CCD exposure. NGC\,1969 is an extended
  low-density cluster. NGC 1971 is also quite elongated} 
\end{figure}

\begin{figure}
\centerline{
\includegraphics[width=8.7cm]{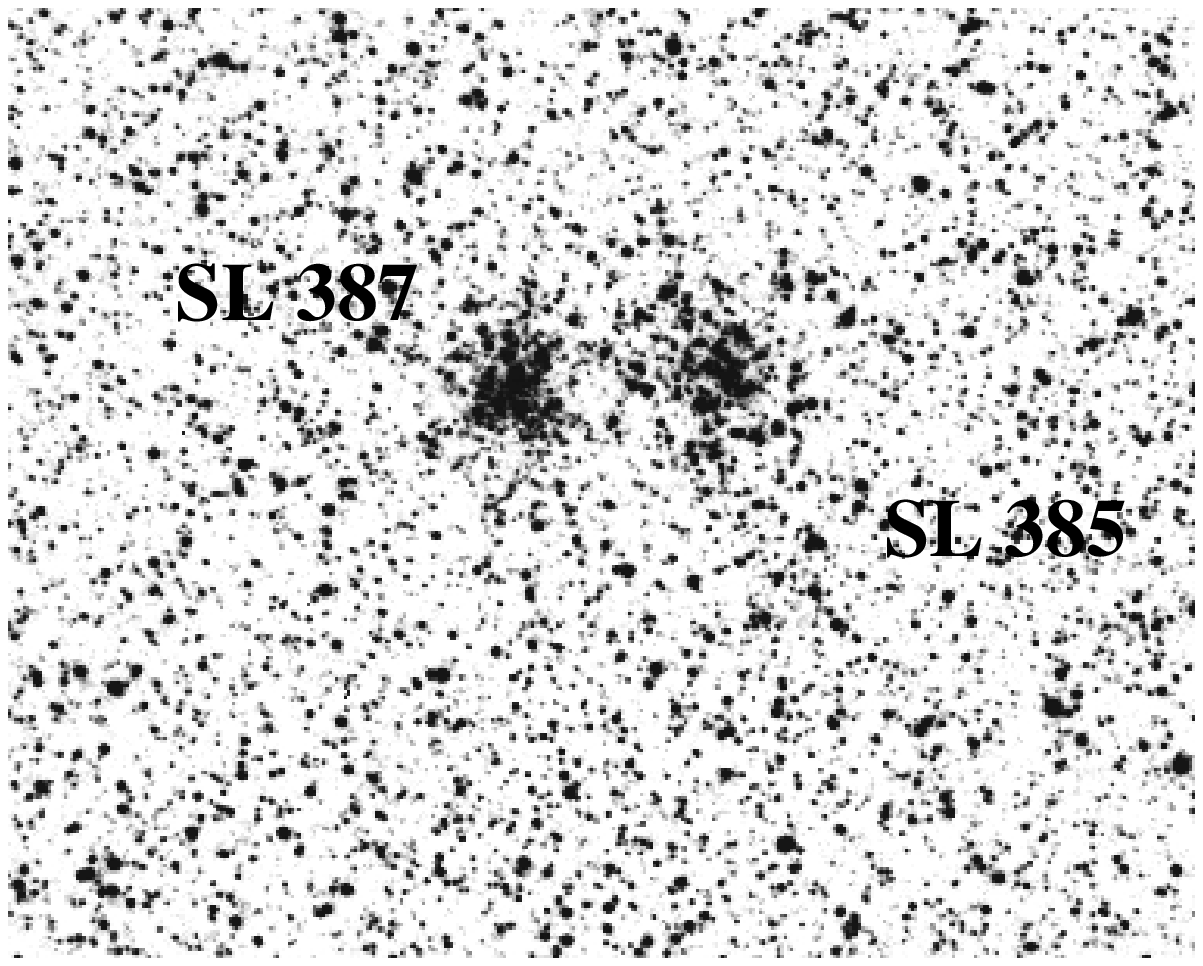}
}
\caption[]{\label{sl385ps} Same as Fig. \ref{n1971ps} but for the cluster pair
  SL\,385 \& SL\,387. This candidate binary cluster is located in the inner
  western part of the LMC bar}
\vspace{0.5cm}
\centerline{
\includegraphics[width=8.7cm]{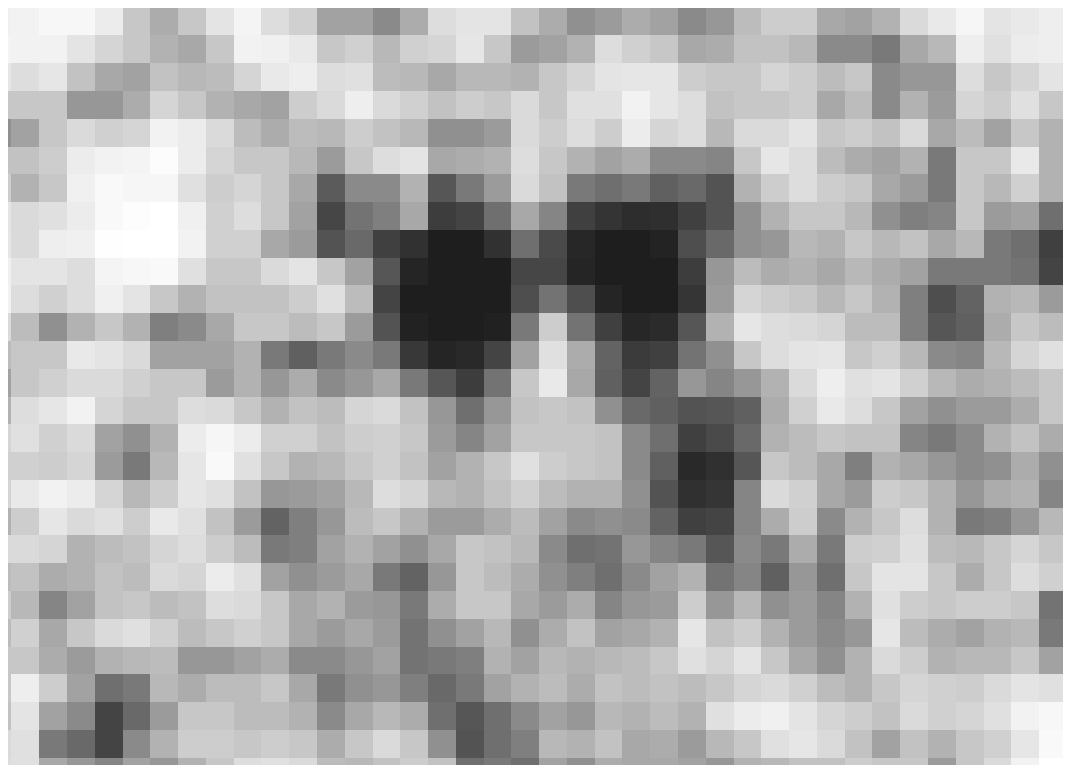}
}
\caption[]{\label{sl385dens} Star density map of the cluster pair SL\,385
  \& SL\,387. The stellar density between the two clusters is enhanced. Note
  the separate enhancement to the southwest of the double cluster}  
\end{figure}

\begin{figure}
\centerline{
\includegraphics[width=8.7cm]{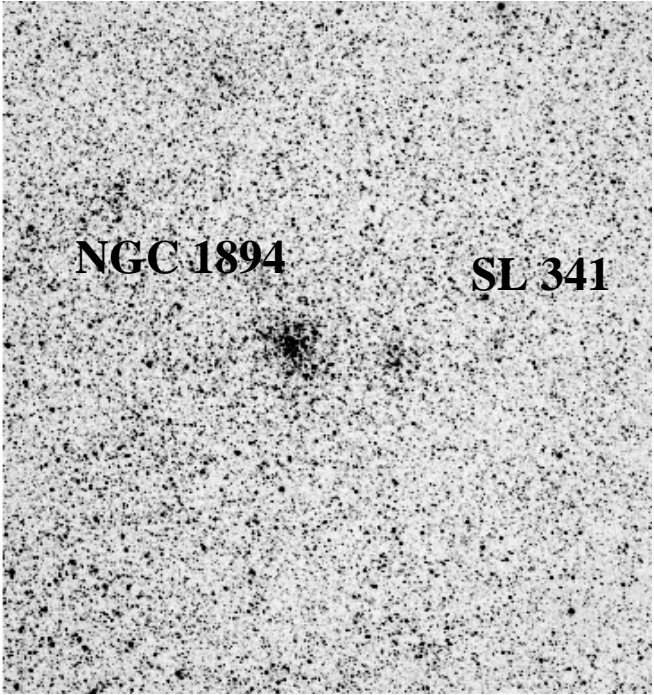}
}
\caption[]{\label{n1894ps} $B$-image of the binary cluster candidate NGC\,1894
  \& SL\,341 taken at the ESO/NTT. The field-of-view is
  $9\farcm2\times8\farcm6$. This double cluster is located at the outer
  south-western rim of the LMC bar} 
\vspace{0.5cm}
\end{figure}
\begin{figure}
\centerline{
\includegraphics[width=8.7cm]{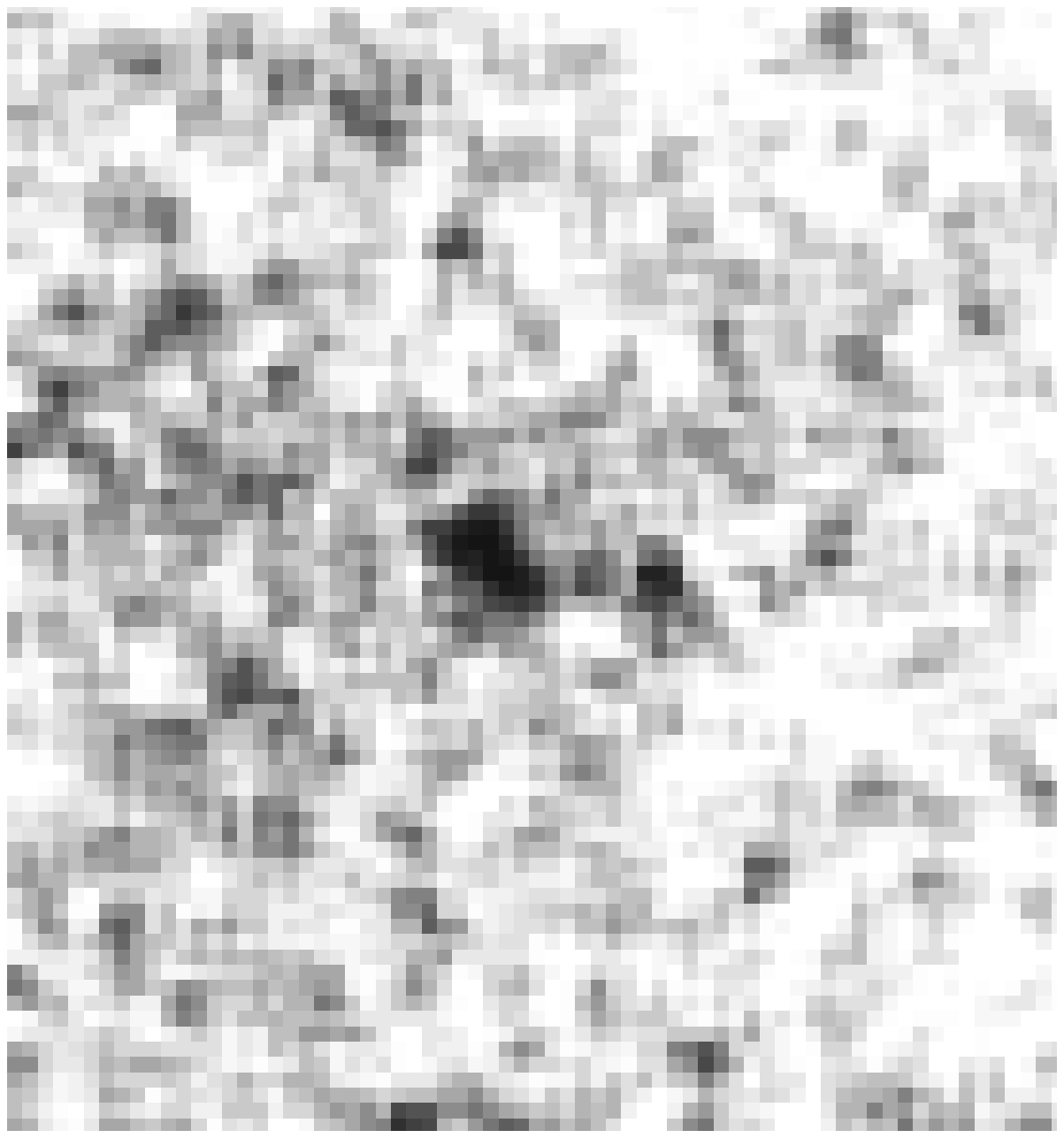}
}
\caption[]{\label{n1894dens} Star density map of NGC\,1894 \&
  SL\,341. NGC\,1894 is more pronounced and has an elliptical shape. Star
  density towards the LMC bar (north-east) is enhanced} 
\end{figure}

We investigate the stellar density distribution in and around the clusters by 
counting the number of stars in square cells of 20 pixels (corresponding to
$6\farcs8$ or 1.7 pc) length. For the images taken with the ESO/NTT we choose
square cells of 26 pixels length. In order to enhance the features of the star
clusters only stars with magnitudes and colours as can be found in
the clusters are included. In this way, a falsification by field stars outside
the selected magnitudes and colours is avoided. To make density structures and
possible signs of interaction better visible we applied a $3\times 3$ average
filter for image smoothing. The procedure is described in  more detail in
Dieball \& Grebel (\cite{dg}). The stellar density maps are shown in
Figs. \ref{n1971dens}, \ref{sl385dens}, and \ref{n1894dens}. 

\subsection{NGC\,1971 \& NGC\,1972:}

Three clusters are covered by the CCD images (see Fig. \ref{n1971ps}) which
may constitute a triple system: the cluster pair NGC\,1971 \& NGC\,1972 
(Bhatia et al. 1991) and to the northwest a third, close cluster is located: 
NGC\,1969. These clusters are situated in the eastern part of the bar of the
LMC. Assuming a distance modulus of 18.5 mag (Westerlund \cite{westerlund})
the projected distances between the clusters are $54\farcs6$ or 13.3 pc
(NGC\,1971 -- NGC\,1972), $76\farcs4$ or 18.6 pc (NGC\,1971 -- NGC\,1969) and
$84\farcs6$ or 20.6 pc (NGC\,1972 -- NGC\,1969).    

For the density plots only stars with $T2\le15.5$ (upper main sequence and
supergiants) or $T2\le19.5$ mag and $M-T2\le0.6$ mag (lower main sequence
without red clump stars) are considered. 
NGC 1969 is a loose extended cluster. It stands out much more clearly on the
stellar density map in Figure 2 than in Fig. 1.
NGC\,1971 is the most prominent
and populous of the three clusters and seems to be somewhat
elongated towards the other two clusters. However, a connection between the
clusters as can be seen e.g. for SL\,538 \& NGC\,2006 (Dieball \& Grebel
\cite{dg}) is not visible from the stellar density map.

\subsection{SL 385 \& SL 387:}

An image of SL\,385 \& SL\,387 is shown in Fig. \ref{sl385ps}. This double 
cluster is located in the inner western part of the LMC bar. The projected
distance between the clusters is $45\farcs6$ corresponding to 11.1 pc.

The selection criterion for the isopleth was $M-T2\le1.8$ and $T2\le19$ mag. 
The stellar density (see Fig. \ref{sl385dens}) between the clusters seems to
be enhanced. The enhancement is between 2 and 3 $\sigma$ above the background
surrounding the binary cluster candidate. To the southwest another separate
enhancement is visible, maybe marking the location of a third, faint cluster.
However, this enhancement is slightly lower than between the cluster pair, but
still between 2 and 3 $\sigma$ above the background.

Leon et al. (\cite{lbv}), whose data for SL\,385 \& SL\,387 go to fainter
magnitudes than ours (see Sect. \ref{ages}), were able to distinguish between
contributions from the two clusters to the possible tidal features around them
by considering the density distribution of stars near the main-sequence
turnoff and along the subgiant branch of both clusters. Their data show that
the cluster-like density enhancement south of SL\,385 may be part of an arced
tidal tail originating from SL\,387. The density extension at the northern
edge of SL\,385 in contrast appears to be part of a tidal feature belonging to
this cluster. These data appear to be the strongest indication of tidal
interactions between two clusters found so far. For a discussion of the ages
see Sect. \ref{ages}.

\subsection{NGC\,1894 \& SL 341:}

An image of the cluster pair NGC\,1894 \& SL 341 is presented in
Fig. \ref{n1894ps}. The two clusters are situated at the south-western rim of
the LMC bar. The projected distance between the clusters is $79\farcs5$
corresponding to 19.3 pc. Thus, this pair shows the largest separation of our
sample.  

Only stars with $16\ge V \le19$ mag and $B-V\le0.5$ mag are included in the
density map (Fig. \ref{n1894dens}). NGC\,1894 is more
pronounced and has an elliptical shape. No significant enhancement between the
clusters can be seen. The star density increases towards the LMC bar in the
north-east and is much lower towards the opposite direction.

\section{Colour Magnitude Diagrams and Isochrone fitting}
\label{ages}

We derived ages for all star clusters by comparing CMDs with isochrones.
The isochrones used are based on the stellar models of the Geneva 
group (Schaerer et al. \cite{smms}). The transformation to the Washington
system was performed by Roberts \& Grebel (\cite{rg}).  

To derive the CMDs for each of the clusters we cut out a circular area
placed around the optical centre of each cluster. Cluster radii were
determined from first stellar density maps which included all stars.   
An investigation of the star density plots suggests that all or at least most
cluster stars are located inside the chosen areas. The cluster CMDs were
compared with the CMDs of the surrounding field and a statistical field star
subtraction was applied to clear the cluster CMDs from contaminating field
stars which  might affect the age determination. 

The CMDs of all clusters and the surrounding fields are plotted in
Figs. \ref{cmdn1971}, \ref{cmdsl385}, \ref{cmdn1894}, and
\ref{cmdfeld}. Overplotted on the CMDs are the best fitting isochrones.  

We only present the $T2$ versus $M-T2$ and $V$ versus $B-V$ CMDs. CMDs with
other colours lead to the same results. 

For all isochrone fits we adopt a distance modulus of 18.5 mag
(Westerlund \cite{westerlund}) and a metallicity of Z = 0.008 
corresponding to $[Fe/H]\approx-0.3$ dex which was found by various authors
for the young field star population of the LMC (Russell \& Bessell \cite{rb},
Luck \& Lambert \cite{ll}, Russell \& Dopita \cite{rd}, Th\'{e}venin \&
Jasniewicz \cite{tj}). We derive reddenings for our selected cluster areas and
compare our values with the reddening maps of Burstein \& Heiles
(\cite{bheiles}) and Schwering \& Israel (\cite{si}).  

Galactic field stars might contaminate our observed area. In order to judge
whether foreground stars might affect our age determination or not we compare 
our CMDs with the number of galactic field stars towards the LMC estimated
by Ratnatunga \& Bahcall (\cite{raba}).
In Table \ref{foreground} we present their counts scaled to our field of
view of $4\farcm3 \times 3\farcm5$ and $9\farcm2\times8\farcm6$, respectively.

\begin{table}
\caption[]{\label{foreground}Number of foreground stars towards the LMC
  calculated from the data of Ratnatunga \& Bahcall (\cite{raba}), scaled to
  our fields of view of $4\farcm3 \times 3\farcm5$ and
  $9\farcm2\times8\farcm6$}  
\begin{tabular}{lccccc}
\hline
                    & \multicolumn{5}{c}{apparent visual magnitude range}\\
colour range        &13-15&15-17&17-19&19-21&21-23 \\
\hline
\multicolumn{6}{c}{field of view: $4\farcm3 \times 3\farcm5$}\\ 
\hline
$B-V < 0.8$       & 0.6 & 1.3 & 1.4 & 2.9 & 2.7\\
$0.8 < B-V < 1.3$ & 0.2 & 1.2 & 2.4 & 1.9 & 3.3\\
$1.3 < B-V$       & 0.0 & 0.3 & 1.7 & 5.7 & 13.1\\
\hline
\multicolumn{6}{c}{field of view: $9\farcm2\times8\farcm6$}\\
\hline 
$B-V < 0.8$       & 3.2 & 6.8 &  7.4 & 15.2 & 14.2\\
$0.8 < B-V < 1.3$ & 1.1 & 6.3 & 12.6 & 10.0 & 17.3\\
$1.3 < B-V$       & 0.0 & 1.6 &  8.9 & 30.0 & 68.9\\
\hline
\end{tabular}
\end{table}

The CMDs and the isochrone-fitting are described in more detail in the
following subsections. 

\subsection{NGC\,1971 \& NGC\,1972 \& NGC\,1969:}
\label{n1971cmd}

\begin{figure*}
\centerline{
\includegraphics[width=\textwidth]{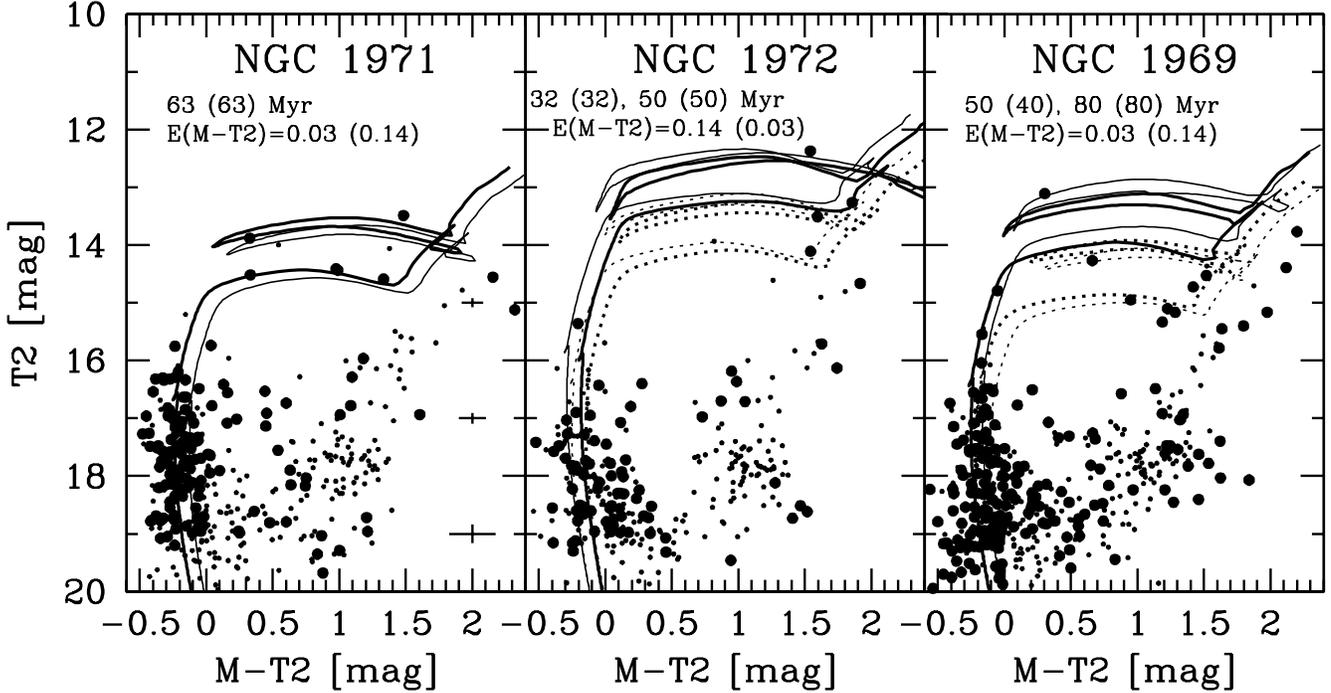}
}
\caption[]{\label{cmdn1971} CMD of the triple cluster NGC\,1971, NGC\,1972,
  and NGC\,1969. Large dots represent stars that remain after field star
  subtraction and thus (statistically) belong to the cluster. Small dots
  represent field stars contaminating the cluster area (radius $30\farcs6$
  for NGC\,1971 and NGC\,1969, $27\farcs2$ for NGC\,1972). The best
  fitting isochrones are overplotted with thick lines. Isochrones of the same
  age, but with a different reddening, are overplotted as thin lines, their
  properties are given in brackets. An isochrone resulting in an age of 63 Myr
  gives a good fit to the main sequence and the supergiants for NGC\,1971. Age
  determination for NGC\,1969 is more difficult since both the 50 Myr and the
  80 Myr isochrones give good fits to either the brightest main sequence stars
  or the red supergiants. NGC\,1972 may be younger with 50 Myr (dotted lines)
  or even 32 Myr (solid lines). 
  Mean error bars are given at the right side in the CMD of
  NGC\,1971. For a more detailed discussion of the isochrone fits see text} 
\end{figure*}

The clusters' CMDs are shown in Fig. \ref{cmdn1971}. These CMDs represent all 
stars located in a circular area with a radius of 90 pixels corresponding to 
$30\farcs6$ or 10.4 pc. NGC\,1972 is the smallest of all three clusters so we 
chose a smaller radius of 80 pixels corresponding to $27\farcs2$ or 9.2 pc. 
The stellar density plot (Fig. \ref{n1971dens}) suggests that all cluster
stars are well within the chosen areas.

Each cluster CMD shows a wide blue main sequence and contains few supergiants.
The width of the main sequence is caused in part by photometric errors
(average seeing $1\farcs3$, see the representative error bars in the CMD of
NGC\,1971) and crowding. The latter one is a major problem in the densely
populated LMC bar. Rotating stars, binaries, and Be stars also lead to a
widening, but will broaden the main sequence only to the red
side. Consequently, we do not fit the isochrones midway through the main
sequence but bluewards from the middle. The small number of supergiants is
expected for poor clusters. Also red clump stars and red giants are 
present in the CMDs. Small dots represent all stars which are located inside 
our selected areas, while large dots denote the stars which remain after
field star subtraction and thus, statistically, belong to the star cluster.
It is evident that red clump stars and red giants, which are mainly
represented by small data points, do not belong to the star clusters, while
most main sequence stars and supergiants are probable cluster members
(represented by large dots). 

After field star subtraction three red and two blue supergiants remain in the
CMD of NGC 1971. They are located within $13\farcs6$ radius except for the
brightest red one which is located within $17\arcsec$ radius. The field has a
few supergiants in the same magnitude range indicating a similar age (see
Sect. \ref{feldcmd}). Scaled to the same area the ratio of cluster to field
supergiants is 5 to 2. We are confident that most if not all of the 5
supergiants belong to the cluster since they are clearly concentrated to the
cluster's centre. An isochrone resulting in an age of 63 Myr and a
reddening of $E_{M-T2}=0.03$ mag gives a very good fit to the 5
supergiants. Also the apparent turnoff of the main sequence is well
represented. Our assumed reddening is somewhat low compared to the 
reddening maps of Burstein \& Heiles (\cite{bheiles}) or Schwering \& Israel
(\cite{si}). Both groups found reddenings of $E_{B-V}\approx0.09$ mag 
corresponding to $E_{M-T2}\approx0.14$ mag (see Grebel \& Roberts \cite{gr},
their Table 5, for the transformation of extinctions in different filter
systems). For comparison we plotted in thin lines also isochrones of the same
age but with a reddening as suggested by Burstein \& Heiles
(\cite{bheiles}). As can be seen, the main sequence, especially its upper
bright part, is not as well represented as with the isochrones of lower
reddening.      

NGC\,1972 is the smaller one of the cluster pair which is also visible from the
sparser main sequence. Four red supergiants remain as likely cluster members
after statistical field star subtraction. The best fitting isochrones result
in slightly younger ages of 32 Myr (solid lines) or 50 Myr (dotted lines). Both
isochrones give a good fit, thus we adopt an age of $40\pm10$ Myr for
NGC\,1972. The isochrones drawn with thick lines are based on a reddening of
$E_{M-T2}=0.14$ mag as suggested by Burstein \& Heiles (\cite{bheiles}), and
give a good fit to the main sequence, both the lower and the brighter
part. Isochrones based on a reddening of $E_{M-T2}=0.03$ mag (as assumed for
NGC\,1971) are plotted with thin lines and fit the blue side of the main
sequence, which might underestimate the photometric error. In the case of
NGC\,1972 a higher reddening seems to give better fits.   

The CMD of NGC\,1969 shows more supergiants remaining after field star
subtraction than the CMDs of the other two clusters. They indicate the young
age of this cluster, but due to their location in a magnitude range between 13
and 15 mag in $T2$ they make an age determination more difficult. Two
isochrones with ages of 50 Myr (solid line) and 80 Myr
(dotted line) are overplotted on the CMD. The 50 Myr isochrone 
gives the best fit to the apparent main sequence turnoff. 
However, the 80 Myr isochrone fits best to the red supergiants. We adopt
an age of $65\pm15$ Myr for this cluster. The lower reddening of
$E_{M-T2}=0.03$ mag (see thick lines) gives a better fit than the value
suggested by Burstein \& Heiles (\cite{bheiles}) (isochrones drawn with thin
lines).    

In all cluster CMDs a population of stars between 16.5 mag and 17 mag in $T2$
can be seen. These data points can be fitted with 200 or 250 Myr
isochrones. However, the distribution of these stars in our field of view is
not concentrated towards the location of the three star clusters, as is, in
contrast, the distribution of the bright main sequence stars (brighter than 17
mag) and bright supergiants. This is an indication of the young age of the
star clusters. Thus we do not plot those isochrones in the CMDs. The few red
stars at $T2\approx16.75$ mag remain in the clusters CMDs due to the imperfect
statistics of the statistical field star subtraction.  

For NGC\,1971 and NGC\,1969 we derived a reddening of $E_{M-T2}=0.03$
mag corresponding to $E_{B-V}=0.02$ mag which is somewhat low compared to the
reddening maps of Burstein \& Heiles (\cite{bheiles}) or Schwering \& Israel
(\cite{si}). Both groups suggest a reddening of $E_{B-V}\approx0.09$ mag
($E_{M-T2}\approx0.14$ mag). 
Considering the photometric and the calibration errors and the uncertainty of
the isochrone fits caused by the width of the main sequence, the error of our
reddening values might be of the order of 0.06 mag, which makes our values
consistent with the findings of  Burstein \& Heiles (\cite{bheiles}) or
Schwering \& Israel (\cite{si}).  
However, our clusters cover small areas with a maximum radius of
$30\farcs6$. Such small scale variations cannot be resolved with either of the
two reddening maps which are derived on scales of $36\arcmin$ (Burstein \&
Heiles \cite{bheiles}) or $48\arcmin$ (Schwering \& Israel \cite{si}),
respectively. For a $1.9^{\circ}\times1.5^{\circ}$ area of the LMC Harris et
al. (\cite{hzt}) provided reddening maps which show variations on smaller
scales. Zaritsky (\cite{zaritsky}) investigated a larger region, including the
field studied by Harris et al. (\cite{hzt}), and found a dependence between
extinction and stellar populations.  
Their fields do not include the LMC bar, so none of our clusters is
included in the region investigated by them.
Another explanation for our low reddening values could be that the star
clusters might be located in front of the LMC bar. 

\subsubsection{Comparison to surface photometry:}

Our study leads to the first ages for NGC\,1969, NGC\,1971, and
NGC\,1972 derived from a comparison between theoretical isochrones and CMDs. 

Bica et al. (\cite{bcd}, \cite{bcdsp}) derived ages for a large sample of LMC
clusters using $UBV$ integrated photometry. Their investigation includes our
binary cluster candidates. However, age determinations based on integrated
colours are less precise than ages derived from CMDs. Geisler et
al. (\cite{gbdcps}) point out that a few bright stars can influence the age
determination based on integrated photometry, making the result dependent on 
the chosen aperture. A comparison between the ages derived by Bica et
al. (\cite{bcdsp}) and our results can be found in Table \ref{photcom}.  
 
According to Bica et al. (\cite{bcdsp}) both NGC 1971 and NGC 1972 have SWB
type II, corresponding to an age of 30 to 70 Myr. NGC\,1969 belongs to SWB
type III which denotes an age between 70 to 200 Myr. Their results are mainly
consistent with our work, but as can be seen from Table \ref{photcom} the age
differences between the components of the cluster pairs cannot be resolved
from integrated photometry. NGC\,1969 is the oldest component of the triple
cluster, which was also found by Bica et al. (\cite{bcdsp}). However, within
the errors all three clusters can be considered as nearly coeval. Indeed, the
integrated colours of NGC\,1969 from Bica et al. (\cite{bcdsp}) place this
cluster in a two colour diagram in the SWB III but near the SWB II
region. Based on our results this cluster could as well be of SWB type II.  

\subsection{SL 385 \& SL 387:}
\label{sl385cmd}

\begin{figure*}
\centerline{
\includegraphics[width=\textwidth]{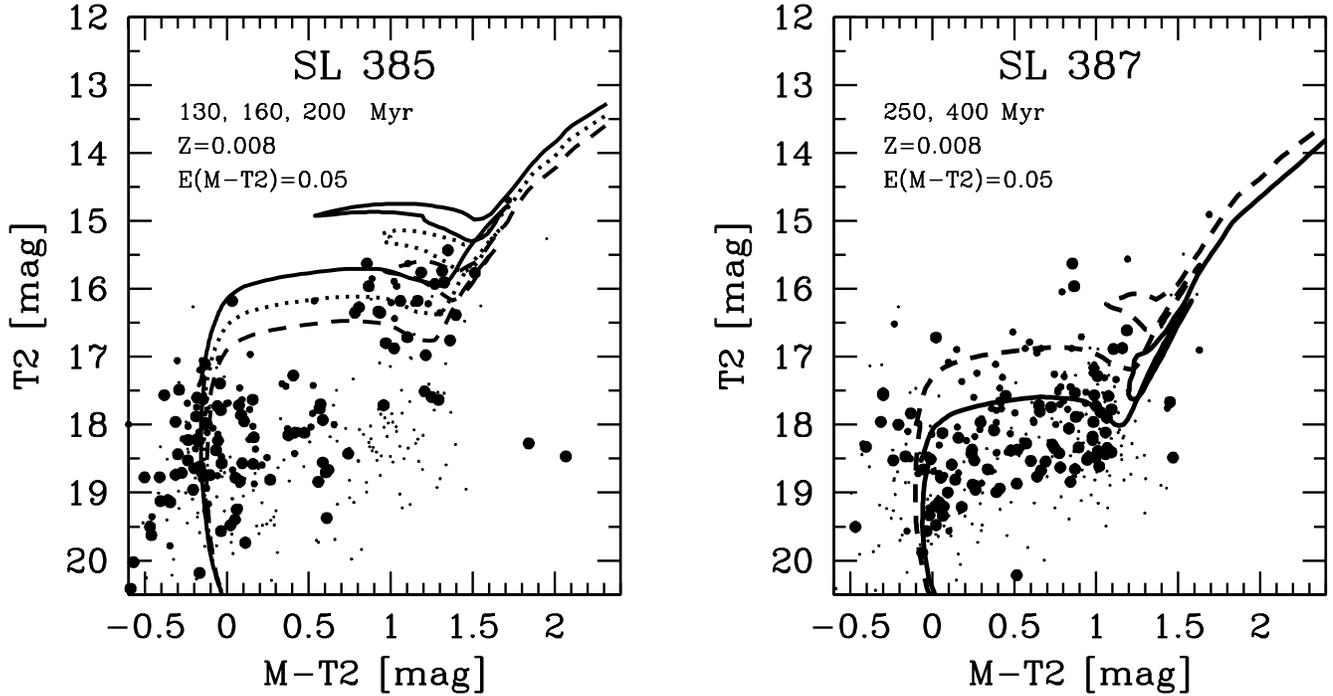}
}
\caption[]{\label{cmdsl385} CMD of the cluster pair SL\,385 and
  SL\,387. Overplotted on the CMDs are the best fitting isochrones. Small and
  large dots denote all stars which can be found in the chosen area of
  $27\farcs2$ radius, only the larger data points remain after field star
  subtraction. Middle size dots in the CMD of SL\,387 represent stars that are
  located in an inner area with $13\farcs6$ radius. Isochrones resulting in an
  age of 130 -- 200 Myr give a good fit to the main sequence turnoff and the
  red supergiants in SL\,385. Sl\,387 seems to be older, note the large amount
  of evolved stars. Fitting the brightest of these stars results in a lower
  age limit of $\ge250$ Myr. For a detailed discussion see text}  
\end{figure*}

Fig. \ref{cmdsl385} shows the CMDs for this binary cluster candidate. 

Following the stellar density distribution we consider stars within a radius
of 80 pixels corresponding to $27\farcs2$ or 9.2 pc as likely members for
SL\,385 and SL\,387. Both clusters appear to have similar sizes, see
Figs. \ref{sl385ps} and \ref{sl385dens}. Again, both large and small dots
represent all stars located inside the chosen region while only the large data
points remain after statistical field star subtraction.  

The CMD of SL\,385 shows a wide main sequence (average seeing $1\farcs4$)
and red giants in a magnitude range between $T2\approx15.5$ to 16.5 mag
and a colour range of $M-T2\approx0.8$ to 1.5 mag. Most of the red  
clump stars vanish after field star subtraction (small dots), indicating the 
younger age of this star cluster. 
Overplotted on the CMD are the best fitting isochrones which represent the 
location of the cluster's red giants as well as the apparent main
sequence turnoff, resulting in an age of 130 Myr (solid line), 160 Myr (dotted
line) and 200 Myr (dashed line). The isochrone resulting in the youngest age
of 130 Myr gives 
the best fit to the main sequence turnoff, but will not fit all supergiants.
The 160 Myr isochrone represents the supergiants well, and gives also a 
reasonably good fit to the main sequence. Also the 200 Myr isochrone would fit
to the supergiants but seems to underestimate the luminosity of the main
sequence turnoff. The luminosity difference between the apparent main
sequence turnoff and the He-core burning giants results in a slight age
difference (130 Myr or 200 Myr, respectively). Since we cannot reject either
the younger or the older isochrone we adopt an age of $170\pm30$ Myr. 

Such a discrepancy between the main sequence turnoff and red giants
luminosities was noticed in the CMDs of various other LMC star clusters. Two
prominent examples are NGC 1866 and  NGC 1850 (see Brocato et
al. \cite{bbcw}, Vallenari et al. \cite{vafcom}, Lattanzio et al. \cite{lvbc},
Brocato et al. \cite{bcp}). Unresolved binary stars which could increase the
luminosity of the main sequence turnoff were proposed as one
explanation. However, as Vallenari et al. (\cite{vafc}) point out, no direct
evidence for a significant population of binary stars could be
found. A good representation of the observed CMDs through simulated ones was
achieved by Lattanzio et al. (\cite{lvbc}), who took both unresolved binaries
and semiconvection into account.

Isochrone fitting to the CMD of SL\,387 is much more difficult. The location
of the main sequence turnoff cannot be determined with confidence, since our
data are not deep enough. The main sequence is affected by crowding and shows
a lot of scatter. The turnoff point is located at $T2\approx18$ mag, at least
one magnitude fainter than in SL\,385 ($T2\approx17$ mag). Dots of three
different sizes are plotted in the CMD. The smallest dots denote the stars
which are rejected by field star subtraction.   
Most red clump stars seem to belong to the field population (smallest
dots) but we see a lot of stars around $M-T2=0.5$ mag between main sequence
and red clump which remain after field star subtraction (medium-sized and
largest dots). 
These stars can also be seen in Vallenari et
al. (\cite{vbc}), their Figs. 2 (left, CMD of SL\,387) and 4 (left, CMD of
SL\,268). It seems likely that the determined colour of these stars is an
effect of the severe crowding in the cluster's centre. The dots of medium size
represent the stars that are located in an inner region with a radius of
$13\farcs6$. As can be seen, the brightest of the stars in question are
situated in the cluster centre. The isochrone which fits best to these stars
has an age of 250 Myr (short dashed line). The large dots in the CMD denote
the cluster stars in the outer annulus between  $13\farcs6$ and $27\farcs2$
radius where crowding should not be a major problem. Fitting an isochrone to
the large dots results in an age of 400 Myr. 

However, neither age is trustworthy. Our data allow us only to set a lower
limit for the age of SL\,387 of $\ge 250$ Myr. 

For all isochrone fits we derived a reddening of $E_{M-T2}=0.05\pm0.01$ mag
corresponding to $E_{B-V}=0.03$ mag which is again lower than the values
derived by Burstein \& Heiles (\cite{bheiles}) or Schwering \& Israel
(\cite{si}) who found $E_{B-V}\approx0.09$ mag (corresponding to
$E_{M-T2}\approx0.14$ mag) (see Sect. \ref{n1971cmd} for the discussion).    

\subsubsection{Comparison to previous age determinations:}

$UBV$ surface photometry for SL\,385 and SL\,387 was obtained by Bica et
al. (\cite{bcdsp}). Based on their integrated magnitudes and colours this
binary cluster candidate consists of two coeval components with SWB type IVa
which corresponds to an age of 200 to 400 Myr. This is not totally consistent
with our findings. Though we derived an age for SL\,387 which agrees with SWB
type IVa, SL\,385 is a younger star cluster with an age of $170\pm30$ Myr which
corresponds to SWB type III. This difference might be explained with the use
of different aperture sizes: Bica et al. (\cite{bcdsp}) used a diaphragm size
of $40\arcsec$ which is that large that also bright stars from the companion
cluster must have been included. We choose a radius of $27\farcs2$ and thus
use a somewhat smaller area from which we derived the CMDs. We cannot rule out
that also our CMDs might be contaminated with stars of the companion cluster,
but these should not be bright stars and thus will not affect our age
determination.  

Vallenari et al. (\cite{vbc}) derived ages for this cluster pair by comparing
isochrones of the Padua group (e.g. Bertelli et al. \cite{bbcfn}) with
CMDs. While we applied a statistical field star subtraction they concentrate
on all stars which can be found in the inner central area with $27\arcsec$
radius. In addition they assume the reddening of $E_{B-V}=0.15$ mag taken from
Schwering \& Israel (\cite{si}). Since small scale variations cannot be
resolved by these reddening maps we derived $E_{M-T1}$ from isochrones fits. 
Vallenari's et al. (\cite{vbc}) findings for SL\,385 are in good agreement with
our results. 
Vallenari's et al.'s (\cite{vbc}) data are $\approx2$ magnitudes deeper than
ours. For SL\,387 they find an age of $\sim 500$ Myr, while we can only derive
a lower limit of 250 Myr. Both data sets are affected by crowding and
Vallenari et al. (\cite{vbc}) discuss the adverse effects of field star
contamination on their age determination. A definitive age determination for
this cluster will have to be postponed until better data are available.

\subsection{NGC\,1894 \& SL\,341:}
\label{n1894cmd}

\begin{figure*}
\centerline{
\includegraphics[width=\textwidth]{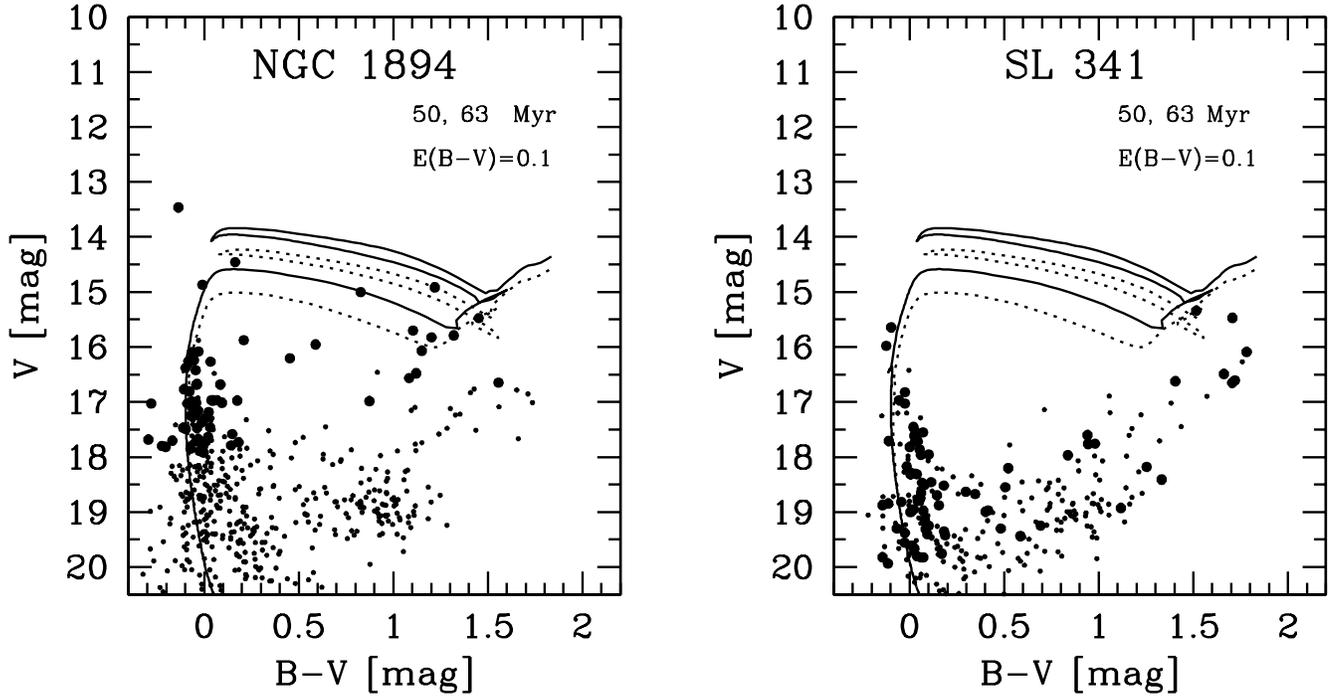}
}
\caption[]{\label{cmdn1894} CMD of the binary cluster candidate NGC\,1894 and
  SL\,341. All stars that are located inside a circular area with a radius of
  $37\farcs8$ (NGC\,1894) or $29\farcs7$ (SL\,341) have been considered (large
  and small dots). Only the large data points remain after field star
  subtraction. As can be seen, the red clump and the red giants belong to the
  field star population. Overplotted on the CMDs are the best fitting
  isochrones. Both the 50 Myr and the 63 Myr isochrone give good fits to the
  clusters' CMDs, indicating that both clusters have similar ages} 
\end{figure*}

The CMDs for the cluster pair NGC\,1894 \& SL\,341 are shown in
Fig. \ref{cmdn1894}. We consider all stars within a radius of 140 pixels
(corresponding to $37\farcs8$ or 9.2 pc) for NGC\,1894, and 110 pixels
(corresponding to $29\farcs7$ or 7.2 pc) for SL\,341, respectively, as likely
cluster members. Again, both large and small dots represent all stars located
inside the chosen region while only the large data points remain after
statistical field star subtraction. 

The CMD of NGC\,1894 shows a pronounced main sequence (average seeing
$1\farcs1$) and some red and blue supergiants around $V\approx14.5$
mag. Nearly all red clump stars and red giants vanish after field star
subtraction (small dots). Fitting isochrones to the apparent main sequence
turnoff at $V\approx16$ mag and to the brightest supergiants results in an age
of 50 Myr (solid line) and 63 Myr (dotted line). Three blue supergiants below
the two isochrones can be seen in the CMD (at $V\approx16$ mag). A 100 Myr
isochrone would fit these three data points but would not fit to the main
sequence turnoff or the brighter cluster giants. Two of the stars in question
are located more than $29\farcs7$ away from the cluster's centre, thus it is
likely that they belong to the surrounding field star population and remain in
the CMD due to somewhat imperfect statistics. Most of the brightest stars are
located inside a radius of $21\farcs6$, thus we are confident that they
belong to the star cluster. 

SL\,341 is a much smaller star cluster. Its CMD shows a bright but sparse main
sequence with a turnoff at $V\approx16$ mag, and only few red
supergiants at $V\approx16$. Most red giants and He-core-burning stars are
rejected by the field star subtraction. Isochrones with ages of 50 Myr (solid
line) and 63 Myr (dotted line) give a good fit to the main sequence turnoff
and to the bluest of the candidate red supergiants.       

We adopt an age of $55\pm5$ Myr for both clusters. Thus, the components of
this pair may be considered coeval.

For all isochrone fits we determined a reddening of $E_{B-V}=0.1$ mag which is
in agreement with the reddening maps from Burstein \& Heiles (\cite{bheiles})
or Schwering \& Israel (\cite{si}).

\subsubsection{Previous age determinations:}

Only for NGC\,1894 a previous age determination can be found in Bica et
al. (\cite{bcdsp}). Using an aperture size of $34\arcsec$ which is slightly
smaller that our adopted radius ($37\farcs8$) Bica et al. (\cite{bcdsp}) found
that NGC\,1894 belongs to SWB type II. This corresponds to an age of 30 -- 70
Myr which is in agreement with our findings. 

It was not attempted previously to determine the age of the companion cluster
SL\,341.  

A summary of previous age determinations derived from the various methods and
ours is presented in Table \ref{photcom}.

\begin{table}
\caption[]{\label{photcom}Comparison of earlier age determinations and our
  results} 
\begin{tabular}{lccc}
\hline
cluster                             & age [Myr]  & reference\\
\hline\hline
                                   &$30 - 70$  &Bica et al. (\cite{bcdsp})\\
\raisebox{1.5ex}[-1.5ex]{NGC\,1971}&$63$       &this work \\\hline
                                   &$30 - 70$  &Bica et al. (\cite{bcdsp})\\ 
\raisebox{1.5ex}[-1.5ex]{NGC\,1972}&$40\pm10$  &this work \\\hline
                                   &$70 - 200$ &Bica et al. (\cite{bcdsp})\\ 
\raisebox{1.5ex}[-1.5ex]{NGC\,1969}&$65\pm15$  &this work \\\hline
                                   &$200 - 400$&Bica et al. (\cite{bcdsp})\\ 
SL\,385                            &$150\pm50$ &Vallenari et al. (\cite{vbc})\\
                                   &$170\pm30$ &this work \\\hline
                                   &$200 - 400$&Bica et al. (\cite{bcdsp})\\ 
SL\,387                            &$500\pm100$&Vallenari et al. (\cite{vbc})\\
                                   &$\ge250$ &this work \\\hline
                                   &$30 - 70$  &Bica et al. (\cite{bcdsp})\\
\raisebox{1.5ex}[-1.5ex]{NGC\,1894}&$55\pm5$   &this work \\\hline
SL\,341                            &$55\pm5$   &this work \\\hline
\end{tabular}
\end{table}

\subsection{The surrounding field populations:}
\label{feldcmd}

\begin{figure*}
\centerline{
\includegraphics[width=\textwidth]{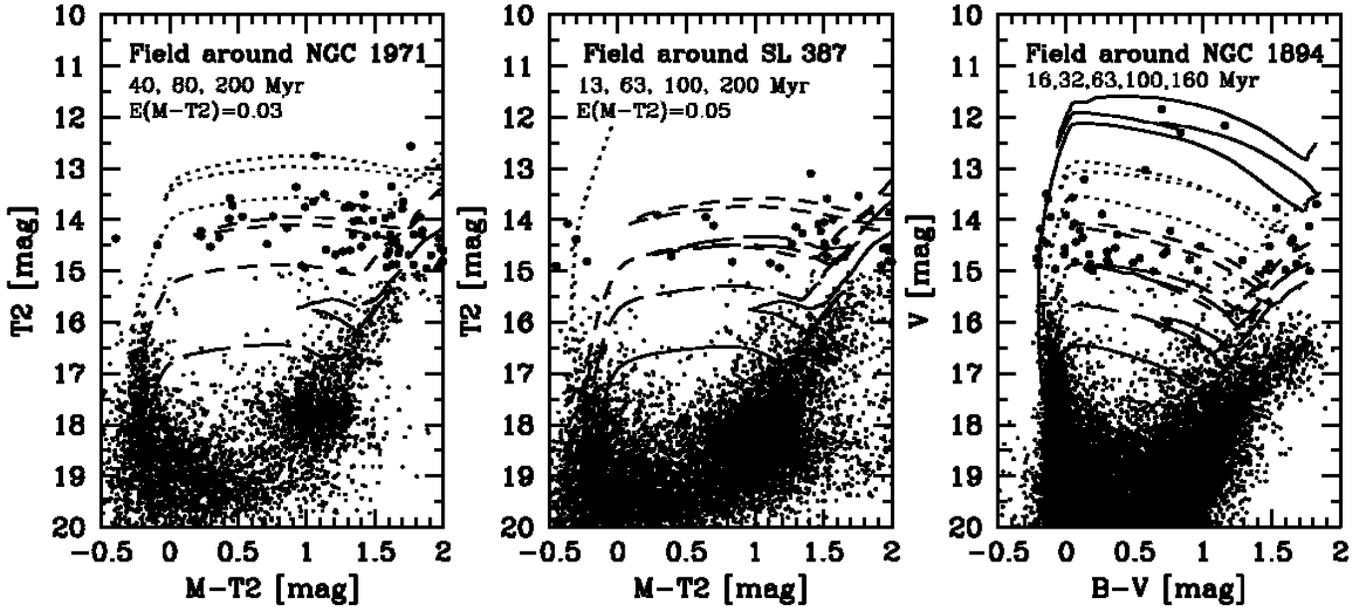}
}
\caption[]{\label{cmdfeld} Left: CMD of the field populations surrounding the 
triple cluster candidate NGC\,1971, NGC\,1972 and NGC\,1969. Stars with
$T2\ge15$ mag are represented as smaller dots to keep the isochrones
recognizable. The field populations comprise a mixture of ages: the blue main
sequence represent the youngest populations that are part of the association
LH\,59, the red giants and the pronounced red clump belong to the intermediate
age populations. Middle: CMD of the field surrounding the cluster pair SL\,385
\& SL\,387. The youngest field population has an age of 13 Myr and thus seems
to be younger than the field surrounding NGC\,1971. Right: CMD of the field
around the double cluster NGC\,1894 \& SL\,341. For all isochrones we adopt a
reddening of $E_{B-V}=0.05$ mag which is lower than the value adopted for the
cluster pair. The brightest stars are fitted by a 16 Myr isochrone. It seems
that the western fields are younger than the field around the triple cluster
candidate}  
\end{figure*}

Both the triple cluster NGC\,1971, NGC\,1972 and NGC\,1969, and the cluster 
pair SL\,385 and SL\,387 are located in the dense field of the LMC bar which 
consists of a mixture of populations of different ages. The cluster pair
NGC\,1894 \& SL\,341 are located at the south-western rim of the bar where the
surrounding field is less dense but shows a density gradient towards the bar
(see Figs. \ref{n1894ps} and \ref{n1894dens}). The CMDs of the eastern and the
western fields are shown in Fig. \ref{cmdfeld}. 

In each CMD a blue main sequence and blue and red supergiants, which represent
the youngest field population, can be seen. The intermediate age population is
represented through red giants and the pronounced red clump. We are not able
to distinguish between distinct young populations, but the few overplotted
isochrones indicate an age range supported by corresponding supergiants.

The eastern field around the three clusters seems to be as young as 40 Myr 
(dotted isochrone) which corresponds to the young age of NGC\,1972. The
youngest field population is part of LH\,59. Also the
80 Myr isochrone (short dashed line, corresponding to the age of NGC\,1969, 
the oldest of the three clusters) is supported by a large amount of red and 
blue supergiants and gives a good fit to the main sequence turnoff. The stellar
density between 80 Myr and 200 Myr (long dashed line) seems to be lower which 
indicates a possible decrease in the field star formation rate. Along the 200 
Myr isochrone and below the stellar density of the He-core burning
giants is increased, indicating enhanced star formation.   

The brightest main sequence stars in the middle CMD of Fig. \ref{cmdfeld}
indicate that the field surrounding the cluster pair SL 385 and SL 387 could
be as young as 13 Myr (dotted isochrone). The 63 Myr (short dashed line) and
the 100 Myr (long dashed line) isochrones are supported by blue and red
supergiants. Below 100 Myr star formation rate seems to be decreased. Star
density along the subgiant branch is again increased along and below the 200
Myr isochrone (solid line). 

The right CMD of Fig. \ref{cmdfeld} presents the field populations around the
coeval clusters NGC\,1894 \& SL\,341. Fitting the brightest main sequence
stars and supergiants results in an age of 16 Myr (solid line). We see a large
amount of supergiants at $V\approx$ 14 -- 15 mag between the 32 Myr (dotted
line) and 63 Myr (short dashed line) isochrone, indicating increased star
formation during which also the cluster pair was formed. Also the 100 Myr
isochrone (long dashed line) is supported by a 
number of giants. Below the 63 Myr isochrone the density along the subgiant
branch is lower, but increases again along and below the 160 Myr isochrone
(bottom solid line).    

From our isochrone fitting we find that the western fields are younger (13 and
16 Myr) than the eastern field (40 Myr). This is consistent with the age
gradient found by Bica et al. (\cite{bcdsp}) for the LMC bar. 

\section{The probability of close encounters}
\label{tidalcapture}

\begin{figure*}
\centerline{
\includegraphics[width=18cm,height=15cm]{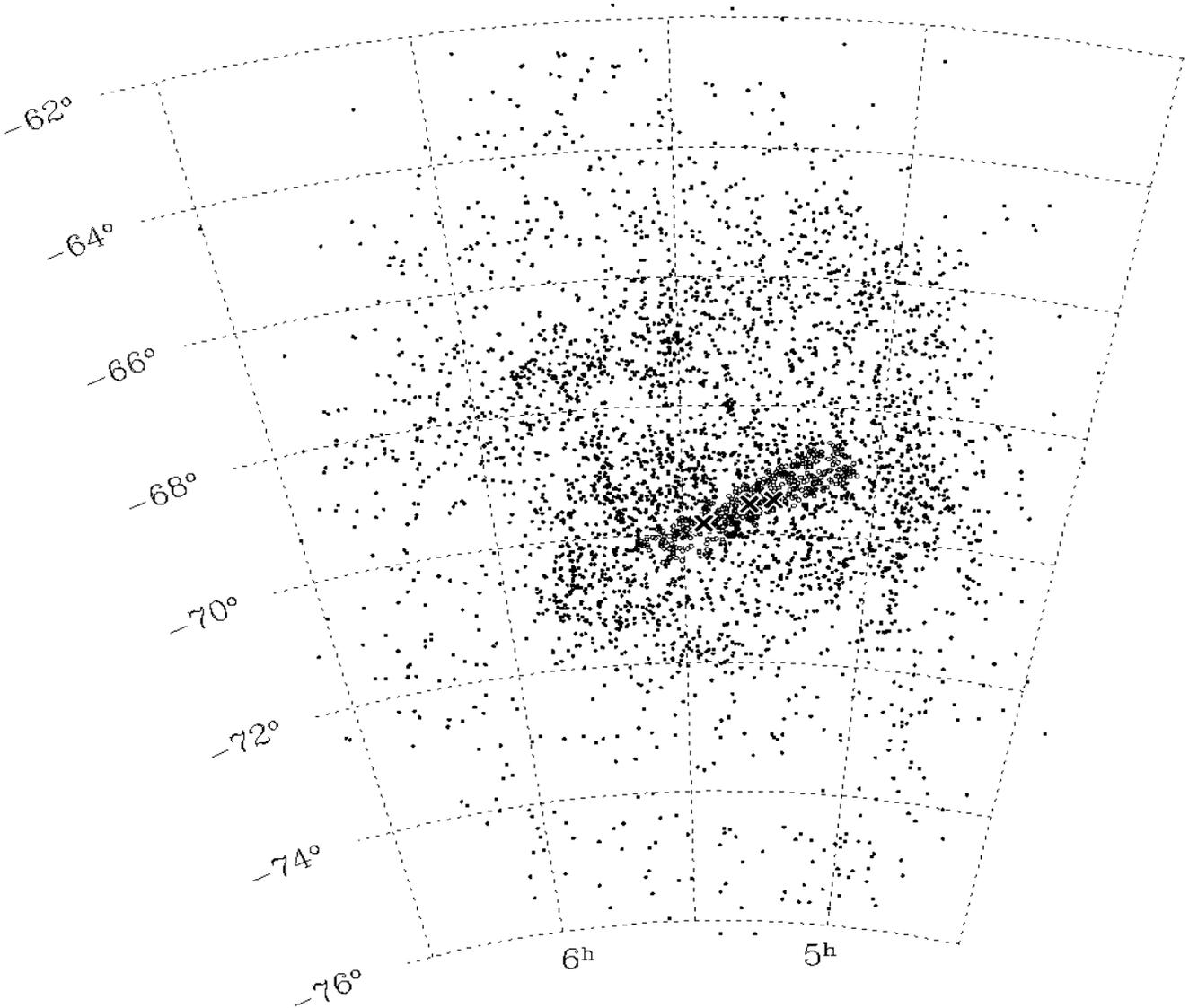}
}
\caption[]{\label{bica} The cluster system of the LMC, data taken from Bica et
  al. (\cite{bscdo}). Open circles denote to the clusters which are located
  in our selected bar region, filled circles represent all star clusters
  situated outside the LMC bar. The black crosses mark the location of the
  three multiple clusters studied in this paper} 
\end{figure*}

The probability of tidal capture of one cluster by another one is small
(Bhatia et al. \cite{brht}), but van den Bergh (\cite{vdbergh}) argues that it
becomes more probable in dwarf galaxies like the Magellanic Clouds with small
velocity dispersion of the cluster systems. In that case the clusters of a
later formed pair do not need to be coeval. Vallenari et al. (\cite{vbc})
estimated a cluster encounter rate of $dN/dt\sim1\cdot10^{-9} \mbox{yr}^{-1}$
in the LMC which makes a capture between young clusters unlikely (for the
definition of $dN/dt$ see below). However, they argue that an interaction
between LMC and SMC, which was suggested in Bica et al. (\cite{bcdsp}), might
also increase the encounter rate. Bica et al. (\cite{bcdsp}) found a
concentration of older clusters (SWB type III -- V, corresponding to ages of
70 -- 2000 Myr) towards the western end of the LMC bar and argue that this may
be the result of an LMC -- SMC interaction $\approx300$ Myr ago. If indeed
such an interaction would increase the cluster encounter rate, we would also
expect a concentration of binary cluster candidates towards the western part
of the bar. We do not see such a concentration in Bhatia et al. (\cite{brht},
their Fig. 5). We re-determine the encounter rate in the LMC bar using the new,
extended catalogue of LMC clusters and associations from Bica et
al. (\cite{bscdo}). Following Lee et al. (\cite{lsm}):
\begin{eqnarray*}
&&\frac{dN}{dt} = \frac{1}{2} \cdot \frac{N-1}{V} \cdot \sigma \cdot v 
\end{eqnarray*}
where $N$ is the number of clusters, $V$ denotes the volume of the galaxy,
$\sigma = \pi R^{2}$ is the geometric cross section of a cluster with radius
$R$, and $v$ is the velocity dispersion of the cluster system of that galaxy.  
For the LMC bar we assume a cylinder 
with a length of $3^{\circ}05\arcmin$ corresponding
to $\approx$ 2700 pc and a width of $40\farcm5$ or 590 pc. Typical cluster
radii are about 10 pc. For the velocity dispersion of the cluster system we
adopt 15 km $\mbox{s}^{-1}$ as quoted in Vallenari et al. (\cite{vbc}). In
Fig. \ref{bica} we plotted all clusters from the Bica et
al. (\cite{bscdo})-catalogue. The clusters which are located in the bar region
are marked with open circles. 484 clusters are situated in our selected
area, leading to an encounter rate of $\frac{dN}{dt} \simeq 2 \cdot
(10^{9}\mbox{yr})^{-1}$.  

Outside of the bar region 3605 star clusters can be found. Assuming a
disk-like shape of the LMC with a radius of $6^{\circ}$ corresponding to 5300
pc and a depth of 400 pc (Hughes et al. \cite{hwr}) we get an encounter rate of
$\frac{dN}{dt} \simeq 3 \cdot (10^{10}\mbox{yr})^{-1}$. Indeed the encounter
rate is increased in the LMC bar by a factor of 10. However, only very few
encounters would result in tidal capture. The probability of tidal capture
will depend strongly on the velocities of the two clusters with respect to
each other, the angle of incidence, whether a prograde or retrograde encounter
takes place, whether sufficiently angular momentum can be transfered, and
whether the clusters are sufficiently massive to survive the encounter. It
seems unlikely that a significant number of young pairs may have formed
through tidal capture.

\section{Summary and conclusions}
\label{summary}

We investigated the stellar densities in and around the triple cluster
NGC 1971, NGC 1972, and NGC 1969. The isopleths do not show connections
between the clusters, but reveal the distortion of NGC\,1971 towards the
other two clusters. NGC\,1969 is much more extended than can be seen from
short exposed optical images. Furthermore we see a clumpy structure on the
stellar density map. Whether this is due to hierarchical stellar clustering
(see text below) or due to background fluctuations cannot be clearly said. 

Comparing isochrones based on the Geneva models (Schae\-rer et al. \cite{smms})
with the CMDs of each star cluster we find the following ages: 
$63\pm10$ Myr for NGC 1971, $40\pm10$ Myr for NGC 1972, and $65\pm15$ Myr
for NGC 1969.

Within the errors all three clusters are similar in age and may have been
formed in the same GMC. The slight age differences suggest that cluster
formation occurred in several episodes, with NGC 1969 as the oldest of the
triple cluster which might have triggered the formation of the other two
clusters, and NGC 1972 as the smallest and youngest object. Similar evidence
for multiple episodes of cluster formation separated by a few tens of Myr have
also been observed in 30 Doradus (Grebel \& Chu \cite{gc}). The 
triple cluster can be seen as an example of hierarchical stellar clustering
(Elmegreen et al. \cite{eepz} and references therein): All three clusters are
embedded in the larger association  LH 59. The youngest field star population
is part of LH 59 and has an age of 40 to 80 Myr which  
corresponds to the ages for the three clusters. The clumpy structure of 
NGC 1969 reveals stellar clustering on smaller scales. Whether the star 
clusters are physically connected or not cannot be clearly decided.
Both NGC 1971 and 1969 show somewhat distorted isopleths which might indicate
a possible interaction, but this is highly uncertain. A thorough radial
velocity study for binary and multiple cluster candidates would be helpful.

Both NGC\,1894 and SL\,341 reveal an elliptical shape on the stellar density
map. A significant enhancement between the clusters was not found. A density
gradient can be seen in Fig. \ref{n1894dens} in direction of the LMC
bar. Whether the elliptical shape of the two clusters is due to interaction or
is a peculiarity that is found for a large amount of single clusters in the
LMC (Goodwin \cite{goodwin}) is unclear.

Fitting isochrones to the CMDs we find that the components of this cluster
pair are coeval within the measurement accuracy with an age of $55\pm5$
Myr. It is likely that both clusters 
have formed from the same GMC. Whether they have or had some interaction cannot
be decided, since most LMC clusters show larger ellipticities than clusters of
the Milky Way. 

The models of both Fujimoto \& Kumai (\cite{fk}) and Theis (\cite{theis}) (see
Sect. \ref{intro}) provide possible formation scenarios for the triple cluster
NGC\,1971, NGC\,1972 \& NGC\,1969 and the cluster pair NGC\,1894 \& SL\,341.  

The star density map of SL\,385 \& SL\,387 shows an enhancement between the
two clusters which connects the components of this binary cluster
candidate. The enhancement is $2 \sigma$ above the mean of the surrounding
field. However, both clusters have a projected distance of $45\farcs6$
corresponding to 11.1 pc and an extension of $\approx27\farcs2$ or 9.2 pc
radius. It is likely that the isopleths of the clusters are overlapping even
if the clusters may have different distances along the line of sight,
e.g. both clusters may appear as a pair only due to chance superposition.
Based on deeper data, Leon et al. (\cite{lbv}) suggest the presence of two
partially overlapping tidal tails distinguishable by their different stellar
populations, which are indicative of a possible encounter between these
two clusters of different ages.
 
Based on isochrone fitting to the CMDs we derived an age of $170\pm30$ Myr for
SL\,385, while only a lower age limit of 250 Myr could be found for SL\,387. 
If we adopt the ages estimated by Vallenari et al. (\cite{vbc}), then the age
difference is $\approx 350$ Myr, far above the maximum life time of
protocluster gas clouds suggested by Fujimoto \& Kumai (\cite{fk}).

We calculated cluster encounter rates for the bar and the disk of the LMC. We
found the cluster encounter rate to be increased by a factor of 10 in the
dense, cluster-rich LMC bar as compared to the disk of the LMC, though the
rate is still very low (2 encounters per Gyr!). Since additional conditions
such as sufficient angular momentum loss must be fulfilled for a tidal capture
of one cluster by another during an encounter, this seems an unlikely
formation scenario for the formation of binary clusters. Obtaining radial
velocities for a large number of stars in cluster pairs, in their potential
tidal tails, and in the surrounding field population would help to evaluate
cluster internal and external dynamics and to rule out or confirm the
occurrence of interactions. 

\acknowledgements

We would like to thank Prof. H. Els\"{a}sser for allocating time
at the MPIA 2.2m-telescope at La Silla during which our data were obtained,
Klaas S. de\,Boer and J\"{o}rg Sanner for a critical reading
of the manuscript, Christian Theis for many discussions about interacting
star clusters, and Hardo M\"{u}ller for his advice in C-programming. This
work was supported by a graduate fellowship of the  German Research Foundation
(Deutsche Forschungsgemeinschaft -- DFG) for AD through the Graduiertenkolleg
`The Magellanic System and Other Dwarf Galaxies' (GRK 118/2-96). EKG
gratefully acknowledges support by NASA through grant HF-01108.01-98A from the
Space Telescope Science Institute, which is operated by the Association of
Universities for Research in Astronomy, Inc., under NASA contract NAS5-26555.

This research has made use of NASA's Astrophysics Data System Abstract
Service and of the SIMBAD database operated at CDS, Strasbourg, France.


\begin{thebibliography}{}
\bibitem[1994]{bbcfn}
Bertelli G., Bressan A., Chiosi C., Fagotto F., Nasi E., 1994, A\&AS,
106, 275
\bibitem[1990]{bhatia}
Bhatia R. K., 1990, PASJ 42, 757
\bibitem[1988]{bm}
Bhatia R. K., McGillivray H. T., 1988, A\&A 203, L5
\bibitem[1988]{bh}
Bhatia R. K., Hatzidimtriou D., 1988, MNRAS 230, 215
\bibitem[1991]{brht}
Bhatia R. K., Read M. A., Hatzidimtriou D., Tritton S., 1991, A\&AS 87, 335
\bibitem[1992]{bcd}
Bica E., Claria J. J., Dottori H., 1992, AJ 103, 1859
\bibitem[1996]{bcdsp}
Bica E., Claria J. J., Dottori H., Santos J. F. C. Jr., Piatti A. E., 1996,
ApJS 102, 57 
\bibitem[1999]{bscdo}
Bica E. L. D., Schmitt H. R., Dutra C. M., Oliveira H. L., 1999, AJ 117, 238
\bibitem[1989]{bbcw}
Brocato E., Buonanno R., Castellani V., Walker A. R., 1989, ApJS 71, 25
\bibitem[1994]{bcp}
Brocato E., Castellani V., Piersimoni A. M., 1994, A\&A 290, 59
\bibitem[1982]{bheiles}
Burstein D., Heiles C., 1982, AJ 87, 1165
297, 37 
\bibitem[1998]{dodb}
de\,Oliveira M.R., Dottori H., Bica E., 1998, MNRAS 295, 921
\bibitem[1998]{dg}
Dieball A., Grebel E., 1998, A\&A 339, 773
\bibitem[1998]{ee}
Efremov Y., Elmegreen B., 1998, MNRAS 299, 588 
\bibitem[1997]{epth}
Ehlerov\'a S., Palou\v s J., Theis C., Hensler G., 1997, A\&A 328, 12
\bibitem[1999]{eepz}
Elmegreen B. G., Efremov Y. N., Pudritz R. E., Zinnecker H., 1999, to be
published in Protostars and Planets IV, eds. V. G. Mannings, A. P. Boss,
S. S. Russell, astro-ph/9903136  
\bibitem[1997]{fk}
Fujimoto M., Kumai Y., 1997, AJ 113, 249
\bibitem[1990]{geisler}
Geisler D., 1990, PASP 102, 344
\bibitem[1997]{gbdcps}
Geisler D., Bica E., Dottori H. et al., 1997, AJ 114, 1920
\bibitem[1997]{goodwin}
Goodwin S. P., 1997, MNRAS 286, L39
\bibitem[2000]{gc}
Grebel E., Chu Y.-H., 2000, AJ 119, 787
\bibitem[1995]{gr}
Grebel E., Roberts W. J., 1995, A\&As 109, 293
\bibitem[1997]{hzt}
Harris J., Zaritsky D., Thompson I., 1997, AJ 114, 1933
\bibitem[1990]{hb}
Hatzidimitriou D., Bhatia R. K., 1990, A\&A 230, 11
\bibitem[1967]{hw}
Hodge P. W., Wright F. W., 1967, The Large Magellanic Cloud, Smithsonian 
Press, Washington
\bibitem[1991]{hwr}
Hughes S. M. G., Wood P. R., Reid N., 1991, AJ 101, 1304
\bibitem[1992]{landolt}
Landolt A. U., 1992, AJ, 104, 340
\bibitem[1991]{lvbc}
Lattanzio J. C., Vallenari A., Bertelli G., Chiosi C., 1991, A\&A 250, 340
\bibitem[1995]{lsm}
Lee S., Schramm D. N., Mathews G. J., 1995, ApJ 449, 616
\bibitem[1999]{lbv}
Leon S., Bergond G., Vallenari A., 1999, A\&A 344, 450
\bibitem[1992]{ll}
Luck R. E., Lambert D. L., 1992, ApJS 79, 303
\bibitem[1970]{lh}
Lucke P. B., Hodge P. W., 1970, AJ 75, 171
\bibitem[1975]{page}
Page T., 1975, in Stars \& Stellar Systems, Vol. 9, p. 541, University of 
Chicago Press, Chicago
\bibitem[1985]{raba}
Ratnatunga K., Bahcall J., 1985, ApJS 59, 63
\bibitem[1994]{rg}
Roberts W. J., Grebel E. K., 1994, AAS 185, 10402 
\bibitem[1989]{rb}
Russell S. C., Bessell M. S., 1989, ApJS 70, 865
\bibitem[1992]{rd}
Russell S. C., Dopita M. A., 1992, ApJ 384, 508
\bibitem[1993]{smms}
Schaerer D., Meynet G., Maeder A., Schaller G., 1993, A\&AS 98, 523
\bibitem[1991]{si}
Schwering  P. B. W., Israel F. P., 1991, A\&A 246, 231
\bibitem[1980]{swb}
Searl L., Wilkinson A., Bagnuolo W., 1980, ApJ 239, 803 
\bibitem[1963]{sl}
Shapley H., Lindsay E. M., 1963, Irish Astron. J. 6, 74
\bibitem[1991]{stetson}
Stetson P. B., 1991, 3rd ESO/ST-ECF Garching - Data Analysis Workshop, eds.
Grosb{\o}l P. J., Warmels R. H., p. 187
\bibitem[1989]{sm}
Sugimoto D., Makino D., 1989, PASJ 41, 991
\bibitem[1997]{teph}
Theis C., Ehlerov\'a S., Palou\v s J., Hensler G., 1997, Proceedings of IAU
Coll. 166 "The Local Bubble and Beyond", Garching, p. 409 
\bibitem[1998]{theis}
Theis C., 1998, in "Dynamics of Galaxies and Galactic Nuclei",
Proc. Ser. I.T.A., Vol. 2, eds. W. Duschl \& Ch. Einsel, p. 223
\bibitem[1992]{tj}
Th\'{e}venin F. , Jasniewicz G., 1992, A\&A 266, 85
\bibitem[1994a]{vafc}
Vallenari A., Aparicio A., Fagotto F., Chiosi C., 1994a, A\&A 284, 424
\bibitem[1994b]{vafcom}
Vallenari A., Aparicio A., Fagotto F., Chiosi C., Ortonlani S., Meylan G.,
1994b, A\&A 284, 447
\bibitem[1998]{vbc}
Vallenari A., Bettoni D., Chiosi C., 1998, A\&A 331, 506
\bibitem[1996]{vdbergh}
van den Bergh S., 1996, ApJ 471, L31
\bibitem[1997]{westerlund}
Westerlund B.E., 1997, `The Magellanic Clouds', Cambridge University Press,
Cambridge, UK 
\bibitem[1999]{zaritsky}
Zaritsky D., 1999, AJ 118, 2824
\end{thebibliography}
\end{document}